\documentclass[11pt]{article}
\usepackage{graphicx, color}
\usepackage{tikz}
\usetikzlibrary{arrows}

\usepackage[sectionbib]{natbib}
\usepackage[colorlinks=true, urlcolor={violet}, citecolor={blue}, linkcolor={red}]{hyperref}
\usepackage{rotating}
\usepackage{verbatim,listings}
\usepackage{fullpage}
\usepackage{amssymb}
\usepackage[reqno]{amsmath}
\usepackage{setspace}
\usepackage{anysize}
\usepackage{lscape}
\usepackage{bm}
\usepackage{booktabs}
\usepackage{changepage}
\usepackage[T1]{fontenc}
\usepackage[utf8]{inputenc}
\usepackage{authblk}
\definecolor{MyGreen}{cmyk}{1.0,0.0,1.0,0.2}

\newcommand{\bt}{\pmb{\theta}}

\usepackage{xspace}
\newcommand{\R}{\textnormal{\sffamily\bfseries R}}\xspace

\begin{document}

\vspace{2cm}

%\textbf{Project To-Do List}
%\begin{itemize}
%\item Refine empirical analysis
%\begin{itemize}
%\item Interpretation: produce dyadic conflict probabilities that we can visualize along with the results
%\item GOF -- discuss
%\item Appendix
%\end{itemize}
%\end{itemize}
%
%\vspace{2cm}
%
%\textbf{Outstanding Questions / Problems}
%\begin{itemize}
%\item Problem with the edgewise shared measure. Need to re-compute in order to ensure that only Type 2 dyads are included here. As it stands, the edgewise shared partner statistic does not distinguish between type 2 and 3 common allies. Thus, the results from model 2 are not meaningful. Models 1, 3, and 4 should still be okay though. 
%\item Can we / how do we distinguish between the prevention of conflict because of agreement on the SQ (no need for conflict) and direct intervention by common allies? Might be done for us by design unless there is a reason to expect that a multilateral alliance indicates \emph{stronger bilateral} agreement on the SQ than a bilateral alliance. So probably, we just need to make this clear in the theory part of the paper. - SJC
%
%\item We use the language of ``mechanisms" in the paper, but we don't really define or test clear, specific mechansims. Do we want to keep this language?  Make the mechanisms clearer?
%\end{itemize}

%\newpage

\title{A Critique of Dyadic Design}

\author[1]{Skyler J. Cranmer\thanks{cranmer.12@osu.edu}}
\author[2]{Bruce A. Desmarais\thanks{bdesmarais@psu.edu}}
\affil[1]{Department of Political Science, The Ohio State University}
\affil[2]{Department of Political Science, The Pennsylvania State University}
\date{}
\maketitle

\begin{center}
\large
\vspace{-1cm}
Forthcoming at {\em International Studies Quarterly}
\end{center}

\begin{abstract}
\noindent Dyadic research designs concern data that comprises interactions among actors. Dyadic approaches unambiguously constitute the most frequent designs employed in the empirical study of international politics, but what do such designs cary with them in terms of theoretical claims and statistical problems? These two issues are closely intertwined.  When testing hypotheses empirically, the statistical model must be a careful operationalization of the theory being tested. Given that the theoretical and statistical cannot be separated, we discuss dyadic research designs from these two perspectives; highlighting model misspecification, erroneous assumptions about independence of events, artificial levels of analysis, and the incoherent treatment of multilateral/multiparty events on the theoretical side and difficult-to-escape challenges to valid inference on the statistical side. 
\end{abstract}

\doublespacing

\section*{\centering{Introduction}}

As social scientists, the objects of our inquiries are social systems. As can be deduced from the subfield's name, International Relations (IR) scholarship focuses on {\em relational} systems. Relational systems are observed as relational data, in which data exhibit two levels of observation -- the actors (i.e., units, or nodes and vertices in network theoretic terminology) and the interactions among actors (i.e., relationships, ties, or edges in network terminology). Relational data\footnote{\doublespacing We should note that relational data are distinct from but potentially related to relational theory. For example, relational social theory -- popularized in voluminous works by scholars such as \citet{tilly2003politics} and \citet{white2008identity} -- takes a network oriented perspective on social relations. Relational data are necessary to test relational theories, but one need not have a relational (network) theory to analyze relational data.} includes both data on single dyads (e.g., a conflict between two states) and data consisting of several actors and simultaneously established ties (e.g., a multi-state defensive alliance). The relationships of interest (e.g., conflict, alliances, trade) are often reduced to observations of whether pairs (i.e. dyads) of nations are incident to domain-specific connections (e.g.,  at war, mutually defensively allied). A research design is dyadic when such dyads are considered as isolated components of the larger relational system. In this essay, we discuss the limitations of dyadic designs for developing explanatory, causal, or predictive models of international relational systems. We show that there exist two conditions under which dyadic designs would be an appropriate approach. Both conditions relate to hyperdyadic dependence; a condition under which the state of dyadic relationships depends upon the states of other dyadic relationships. The first condition is simply the assumption that there are no hyperdyadic dependencies (i.e., that dyads do not depend upon each other). The second condition is an absence of interest in hyperdyadic dynamics paired with an assumption of no confounding of systemic patterns and covariates of interest. If neither of these conditions hold, it is our position that strictly dyadic designs are inappropriate and will often lead to faulty inferences. 

% paragraph on the rise of dyadic design in IR from a historical perspective
In the past several decades, the study of interstate conflict has adopted a heavy focus on dyads as a unit of analysis (the dyadic literature is \emph{far} too voluminous to begin citing here, but consider just a few prominent examples that explicitly discuss dyadic design from author's whose names begin with ``B'': \citet{BdmLalman1992, Bremer1992, BremerReganClark2003, Bennett2004}). A great deal of work has focused on dyadic attributes as explanations for peace and violence. Theories in this vein consider geographical mechanisms, like whether the states in the dyad are contiguous, as well as relational mechanisms, like whether the states share alliances or memberships in intergovernmental organizations, or economic and diplomatic relations within the dyad, as predictors of the probability of conflict. Much has been made of joint democracy within the dyad \citep{Oneal1999, MaozRussett1993,Ellis2010}, as well as differences in power and capabilities, also frequently feature in these studies. Despite this wide range of explanations, however, these theories all share an emphasis on the influence of direct ties between states. Recent work has shown that a focus on dyads is both theoretically incomplete and empirically problematic \citep{Cranmer2011a, Cranmer2012b, Cranmer2012a, manger2012hierarchy, Ward:2007, Warren2010,Ward2011,Dorff2013}. The focus on dyads overlooks extra-dyadic and systemic effects that might yield interesting findings. Furthermore, a dyadic analysis necessarily ignores the fact that states in the international system are embedded in a dense web of economic, political, and social ties. 

%\sjc{The problem is widespread:} \red{Try to gather some data on how frequent dyadic designs are in IR. A simple jstor search for ``dyad'' within journals that publish IR  will capture a few other things, like our complaining about dyads, but should be mostly right}

Lastly, we note that our discussion is limited to a critique of dyadic design. The reader may be dissatisfied with the presentation of problems absent solutions, but our essay is restricted to the subject of dyadic design. We attempt to stay strictly on that topic, though we do allude briefly to some solutions to some of the problems we discuss. Suffice it to say that alternative methodologies exist and are well-implemented in software that can solve most of the problems we highlight here, but a detailed discussion of these techniques is beyond the scope of the present essay. %See CRANMER2015 for an overview of three of the most common approaches to statistical inference on relational data. 

\section*{\centering{The Difference Between the Study of Relationships and Dyadic Design}}

% relationships != dyadic design
The first thing we wish to highlight is that a dyadic design is different from an interest in relationships. Indeed, the dyad is the simplest level of measurement at which a relationship can be recorded. Sets of relationships are usually studied as dyads and dyadic relationships are an item of interest even in network scientific approaches to relational systems -- which include, but are not limited to, dyadic designs. An interest in relationships is fundamental to the study of international relations. 

% Define dyadic design clearly
What does a dyadic design mean? Dyadic design refers to the empirical design commonly applied to the study of international politics in which dyads, either directed or undirected, are the units of analysis and a regression model is applied with dyadic measurements as the outcome variable. One key feature of a truly dyadic design is that the dyadic outcome observations are not used to explain each other. We use the common notation of $y_{ij}$ as the outcome variable observation for the dyad including state $i$ and state $j$ (undirected) or $i$ to $j$ (directed). In a dyadic design, the covariates used in predicting $y_{ij}$ do not involve $y_{kl}$ $\forall~\{i,j,k,l\}$ (i.e., the other dyads in the system).\footnote{\doublespacing Readers may note that this definition excludes logistic regression models that involve network statistics on the right hand side. That is intentional. Such models constitute conditionally specified \cite{Chen2010} exponential family random graph models, and the forms of dependence they imply are more than dyadic.}  For example, the ``standard'' model of conflict applies such a design.  In these models, the outcome variable is the presence of conflict, measured dyadic ally, with potential predictors including (but not limited to) joint democracy, trade dependence, joint IGO membership, and CINC ratio\footnote{\doublespacing We consider this somewhat generic model rather than pick on any one particular scholar who has published an empirical analysis with a dyadic design.}.   Such a model is then analyzed using logistic regression either on the set of all dyads or only politically relevant dyads. In this discussion, we voice criticism of two problematic aspects of such dyadic designs: \emph{dyadic theory}, the (potential) pitfalls of thinking dyadically and the misspecification  of theory that can arise, and \emph{dyadic analysis}, the practice of analyzing dyadic data with regression-style techniques. We will show that models such as the one described above are theoretically limited to the point of misspecification and sufficiently problematic empirically that one cannot have confidence in inferences drawn from such a model.

% paragraph on the study of relationships from a non-dyadic perspective
An interest in relationships need not translate into a dyadic design such as that above or discussed in detail below. A variety of approaches facilitate the study of relationships without the restrictions of dyadic designs. In our own research, we have come to view relational phenomena in international politics as network phenomena; a perspective that, we believe, provides a sound treatment of relational data as forming interconnected relational systems. Relational hypotheses from such a perspective make use of dyads: theory may suggest certain processes occurring in the dyadic relations between two states and complex networks are the product of many dyadic relationships. But a perspective that leverages information from outside the dyad and considers more than dyadic phenomena as part of the model necessarily breaks from the tradition of dyadic design in IR.

\section*{\centering{Limitations of Dyadic Theory}}
In this section, we lay out a critique of dyadic design from a theoretical perspective. Not all theories tested with a dyadic design face the problems we mention below. For most political scientists, theory is more easily malleable than empirical models and many examples may be found in which the theory is free of the shortcomings we discuss, but the empirical analysis is not. %, we will even point out some of these works ourselves. 
That said, we do believe that theoretical problems discussed below are endemic to the study of international politics and represent a serious challenge to the development of sound, consistent, and complete theoretical explanations for a range of important international phenomena. Thusly, we propose the following as a criticism of what we see as trends in the literature rather than absolute and intractable problems of theory.  We also note that some of the theoretical critiques discussed below are implied by the statistical model (per definition a careful operationalization of theory) if not made explicit by the researcher. 

\subsection*{\centering{\textnormal{\textit{The difference between a model and a hypothesis}}}}
A preface to our theoretical critique of dyadic design is that we must first be clear on the difference between a hypothesis about a process and a model of the same process. Both are deduced from theory. A hypothesis, at least as commonly used in international relations, is a simple statement of how, based on an underlying causal theory, two processes are expected to covary. For example, one expects that as trade between a pair of states increases, the probability of those same states going to war with one another is diminished. Such hypotheses are usually the primary interests of scholars who develop research designs, including dyadic designs, for relational data. Over time and multiple studies, we build empirical support for such hypotheses and thus come to believe that they and their underlying causal theory are a step in the right direction with respect to understanding the process by which nature generates the phenomena of interest. 

A model is a comprehensive specification of the entire data generating process% that reflects all hypotheses related to the underlying data generating processes
. If correctly specified, a model will generate artificial datasets that stochastically approximate observed data. In other words, a model includes all relevant predictors and other generative features, which incorporate the entire range of phenomena likely to cause the outcome. When conducting model-based inference, in which each hypothesis is evaluated with reference to a parameter in a statistical model of observational data, a comprehensive model is needed to test hypotheses (i.e., as compared to design-based inference, in which the researcher exercises experimental control of data generation). Correctly specifying the model eliminates the bias caused by omitted variables that are casually related to the outcome. The sole exception to this, and the reason we say ```near complete'' is because a model can be a less-than-complete description of the entire data generating process in the event that those omitted factors are orthogonal to the predictors. In other words, in order to leave certain factors out of a model, we must have reason, at least theoretically if not empirically, to think that these omitted factors have no systematic relationship to any of the predictors. In such a case, it is only necessary to use a robust covariance matrix in parametric inference or nonparametric inference procedures such as the jackknife or bootstrap.

%So a model is a holistic description of the data generating process, that expands beyond the hypotheses specifically devised from the causal theory. This holistic model is also derived from theory. Often this will mean the inclusion of variables that are not necessarily interesting to the researcher, but seem like likely causes of the outcome, whether consistent with the theory underlying the hypothesis or being drawn from an alternative theoretical framework. 

Our point here is to emphasize the fact that, just as important as what is included in the theory, is what is omitted from the theory. Statistical models, from which we find either support or a lack of support for our theories, are careful operationalizations of theory, so it is worthwhile to consider some things that are assumed orthogonal (and thus not modeled) in dyadic designs from a theoretical perspective. 
 
\subsection*{\centering{\textnormal{\textit{Assumption: dyads exist in isolation}}}}
A fundamental assumption of dyadic analysis is that dyads exist in isolation from anything else going on in the world, with respect to the outcome variable, during the period of observation for the dyad. This may not be an assumption the analyst wishes to make, but it is strongly implied by the application of a dyadic quantitative design, such as a dyadic regression model. Fundamental to the regression model, as discussed from a statistical perspective below, is the assumption that observations are conditionally independent from one another, that conditioning being on the included predictors. Translating this into the terms of our example model from above, this implies that, conditional on things like shared regime, trade dependence, and CINC ratio, the dyad of states $i$ and $j$ is \emph{entirely unrelated} to anything else happening in the world. This includes $i$'s relationships with other states (e.g. the dyads $ik$, $il$,\ldots) and all other dyads (e.g dyads $ab$, $cd$, \ldots). Let us consider a few basic logical tests of this assumption. 

Consider the first type, involving the multiple relations of one state, first. A more detailed examination of an example given by \citet{Cranmer2012a} will illustrate the point nicely. In 1939 a new dyadic conflict connection was formed between the UK and Germany. Let us denote this dyad $bg$ ($b$ for Britain). Yet there was, at least, one other dyad formed that year: by the German invasion of Poland. Let us call this dyad $gp$. Dyadic design requires (and thus assumes) that the $bg$ and $gp$ dyads are\emph{entirely} unrelated to each other, conditional on things like an indicator for common regime type and CINC ratios. Were one to proffer such an explanation for these ties at a professional meeting of conflict scholars, one would be laughed out of the room. Yet this is precisely what is implied by a dyadic design applied to the problem of international conflict. Needless to say, such an assumption has been tacitly made whenever dyadic analyses have been published and much of our understanding of international phenomena ranging from war to trade is based such tacit theoretical claims. It is also worth noting here that the fact that two events are separate by no means implies that they are independent. Indeed, mutual exclusivity is a strong form of dependence. So, just because the decisions by the German and British governments respectively to wage war on Poland and Germany, also respectively, does not in the least imply that they were independent. The British government gave the invasion of Poland as its primary reason for war with Germany and, absent such an invasion, the war may well not have occurred.

We can use the domain of defensive alliance formation to further probe the validity of the assumption that dyads $ij$ are independent of other dyads $ab$. Let us consider the First World War. Is it reasonable and consistent with both theory and history to assume that, say the British alliance with France (dyad $bf$) is completely independent of the alliance between Wilhelmen Germany and Austria-Hungary (the dyad $ga$)? Given that the one alliance was formed in order to balance the other, it seems misguided to search primarily for explanatory power in dyadic covariates such as regime similarity and CINC ratios. 

%% BD, I know that the following paragraph turns out to be the case in many dyadic studies, but its not true of them all, and its not difficult to map state covariates to dyads or system covariates to dyads if the data are temporal.

%The assumption of independence inherent in dyadic design is even stronger than what is implied above. A dyadic design also implies that state and system level factors are not part of the data generating process, or are at least orthogonal to any dyadic predictors included in the model. Let us consider each of these implications in turn, and we will do so again in the context of an example. First, consider the relevance of state attributes, even if the outcome of ultimate interest is relational. A dyadic design cannot naturally accommodate Would it be fair to say that the international politics of Saudi Arabia were unrelated to its production of oil? In other words, would we expect the oil production of Saudi Arabia to be unrelated to those factors that might predict war, alliances, trade, or any number of other relations that scholars of international politics study? Clearly not in our opinions. 

And what of phenomena occurring in the larger neighborhood of the focal dyad (and possibly involving it)? In a study of international alliances, \cite{Cranmer2012b} theorize that a dyadic alliance is more attractive to the prospective allies when that alliance is embedded within a dense clique of states that are all allied to each other. This is an application of the common network theory that triadic bonding reinforces dyadic trust. Both \cite{Cranmer2012b} and \cite{Cranmer2012a} show that triadic closure is a relational process that is key to accurately predicting the ultra-dense cliques that characterize the defense alliance network. Is it reasonable to think of this phenomena, which occurs at the triadic (or higher) rather than the dyadic level, as either not being a meaningful part of the data generating process, or as being completely orthogonal to the dyadic predictors included in a given model? Such assumptions would not be tolerated if directed at a strongly performing control variable, such as same regime type or relative military capabilities, but that assumption is made any time a dyadic research design is applied. 

To argue the validity of the dyadic design, one may object that the variety of hyper-dyadic dependencies just discussed, while maybe not unimportant for the data generating process, are orthogonal to the dyadic predictors included in the model. Such an objection is to take the attitude that, though strictly dyadic models are substantively misspecified, hypothesis tests derived from the dyadic model are robust to this particular form of misspecification. Such an assumption, which boils down to one of orthogonality of all predictors to the higher order system factors, is rather bold. It is the case that, in the event of orthogonality, an empirical analysis is not biased by the omission of such factors and thus model misspecification is less consequential -- possibly only affecting covariance estimates. Just as it is rarely justified to assume that important omitted covariates are orthogonal to those one has included, there is no sound scientific basis upon which to presume that covariates are orthogonal to the effects of system structure on dyadic outcomes.

%Certainly, there have been cases of theories 
\subsection*{\centering{\textnormal{\textit{Artificial Levels of Analysis}}}}
 %Theoretical Myopia and the drunkard's search

%\sjc{Should we strike this section for the same reason as the paragraph some ways above? Namely, that is is possible to link state level predictors to dyadic analyses, just the people rarely do it.}

% The second paragraph creates a nice bridge between the levels discussion and the broader point of the paper.

The so-called ``levels of analysis'' in international relations -- state, dyad, and system (see e.g. \citet{Mitchell1999} and \citet{Bennett2004})--  are artificial and are products of our measures and the units of analysis on which we focus in common regression models. This is not a problem unique to dyadic design, but dyadic design is one incarnation of the problem. But let us consider this problem in the context of dyadic design. The critical theoretical question one must ask is: though our outcome of interest is relational (dyadic), what makes us think that \emph{all} of the relevant predictors of this phenomena are also relational and concern \emph{only} the relations between the two states in the focal dyad? In other words, a focus on independent dyads neglects the fact that what happens at other levels of aggregation of dyads (e.g., state, entire system/network, region, triads) has implications for the dyadic outcomes of which the higher level outcomes are comprised. For example, \cite{Cranmer2014son} show that there are popularity (i.e., pile-on) effects in the network of international economic sanctions. When one state sanctions state A, other states then become more likely to sanction state A. Theory specified at the dyadic level implies that no such higher order (i.e., hyperdyadic) patterns in tie formation exist. 

%But suppose one had a genuinely dyadic theory: a process between $i$ and $j$ is thought to affect conflict between those two states. Would this not be a case where dyadic design is truly appropriate? We argue generally not for the following reasons. First, for the results to be unbiased, one must assume that no processes above of below the dyadic level affect the dyadic outcome. 

We argue that the focus on any one level of analysis, including dyadic analysis which can be said to be the most common of the levels of analysis, is either (a) a very bold claim about the data generating process, or (b) an example of theoretical myopia driven by the availability of data (e.g. mostly dyadic) and the data formatting constraints associated with conventional statistical models (e.g. traditional regression). Most would find (a) highly unrealistic substantively, so we suspect that much of the focus on dyads is driven by (b). Yet this is an example of a drunkard's search; searching for one's keys under the lamplight because that is where one can see rather than because that is where one believes them to have been misplaced. If (b) occurs, this is obviously bad for the scientific process: our theories should not be driven by available techniques and data, but rather data should be gathered and statistical techniques developed, if necessary, to follow theory.

%\subsection{False ``dyadization'' of state characteristics}
%\red{Do we even want to talk about this?}
\subsection*{\centering{\textnormal{\textit{Incoherent treatment of multilateral/multiparty events}}}}
Thus far we have presented dyadic outcomes as being interdependent -- influencing each other. For example, \cite{Cranmer2011a} demonstrate that the international conflict network is intransitive in that two states -- A and B -- in conflict with a third state C -- are statistically unlikely to be at war with each other. This is an interdependence result regarding individual dyadic outcomes, including the absence of a tie (i.e. a conflict). The key feature of the network modeling approach is that it is not required that the analyst assume these separate dyadic outcomes are {\em independent} dyadic outcomes. This leads to our last theoretical critique of the dyadic independence approach. Even though it may not cause bias from a statistical perspective, dyadic design is problematic with respect to the sensible treatment of multilateral and/or multiparty events. The nature of this problem should be obvious from our above discussion of WWII. Was the German invasion of Poland a different war from the British war with Germany? Was the American war with Germany different from the Canadian? There were certainly differences in the way these countries participated in the war; in their strategies, their forces brought to bear and so forth, but they were not different wars.  \citet{Cranmer2012a} point out that, dyadically, WWI and WWII constitute over 500 different wars (completely independent remember, conditional on regime similarity and CINC ratios if we are applying a dyadic design) and more than half of the wars in the 20th Century. 

The lack of face validity to this treatment of two wars should be cause for pause from analysts seeking to model any relational phenomena that is, or has the potential to be, multilateral. Multilateral events, of course, consist of several dyadic relations that presumably could dissolve while the others stay in place, but multilateral events often arise due to complex interdependence among actors and dyadic relations. A dyadic design simply cannot accommodate multilateral data without imposing strong assumptions about the independence of dyadic events; assumptions that at worst bias statistical models and risk faulty inference, or at best result in a grossly unrealistic view of the data and their corresponding generating process (i.e., that the alignment of multilateral relations is purely coincidental). 

A variety of solutions to this problem have been posited in the literature. For example, \cite{Poast:2009} proposed a $k$-adic rather than dyadic design. A $k$-adic tie exists among $k$ states if all $k$ states are tied to each other (e.g., as in a multilateral alliance).  In our own research, we have adopted a network perspective that treats the entire complex network of ties as an observation, allowing for near arbitrary order dependencies to exist in the data. By explicitly modeling the interdependence among relationships, network models can be tuned to predict the alignment of $k$-adic relations (with heterogeneity in $k$), which codify as multilateral relational outcomes. For example, as discussed above, network models of defense alliance formation predict strong triadic closure effects, which leads to a network of moderate to high-order $k$-ads \citep{Cranmer2012a,Cranmer2012b}.

\section*{\centering{Limitations of Dyadic Methods (regression)}}

In this section we discuss the limitations of dyadic analysis, which we take to mean regression analysis on dyadic data, from a statistical perspective.  We should be clear that we are not making a case against the development and specification of hypotheses based on dyadic covariates. Dyadic covariates represent an essential class of terms that should be included in hyperdyadic models. Rather, our critique of dyadic regression applies to the use of regression models for dyadic dependent variables to estimate the effects of exogenous covariates without accounting for hyperdyadic dependence. 

\subsection*{\centering{\textnormal{\textit{Problematic inference in realistic applications}}}}
The major concern we have with dyadic design and analysis, more pressing even than the theoretical limitations discussed above, is the fact that dyadic designs will produce biased estimates and incorrect inferences in hypothesis testing in most applications. The conditions necessary for a dyadic design to be unbiased are essentially the same as those laid out above in the section discussing theoretical limitations. It must be the case that all dyadic observations are conditionally independent from one another. In other words, conditional on the right hand side of the regression, it must be the case that the relationships among states do not affect other relationships among states (including the other relationships of members of a given focal dyad), state or substate attributes and processes are not allowed to affect the outcome in any way that might be related to the included dyadic predictors, and that neither proximate (e.g. neighborhood) or system level effects may affect the outcome and be related in any way to the dyadic predictors. Given a century or more of development of the modern globalized economy, the activity of highly developed institutions of international law and governance, and the rapid global flows of information; it is hard to imagine any context in international relations in which an across-the-board assumption of dyadic independence would be tenable. 

To see how misspecification manifests, let us consider the matter of statistical independence. The formal definition of statistical independence is: 
\[
p(a \cap b) = p(a)p(b). 
\]
In other words, the joint probability of some event $a$ and some other event $b$ is the product of their marginal probabilities. The reasoning behind this definition is made more clear when we consider that, from the basic laws of probability, we have: 
\[
p(a | b ) = \frac{p(a \cap b)}{p(b)},
\]
which can be rearranged to show that 
\[
p(a \cap b) = p( a | b) p(b).
\]
The above equation will only reduce to $p(a \cap b) = p(a)p(b)$ in the event that $a$ and $b$ are independent. The extension of the above to the case of conditional independence is simple: if independence is conditional on some third event $c$, we have $p(a \cap b | c) = p(a | c) p(b | c)$. 

Statistical independence is fundamental and inexorable from regression-based statistical models. For illustration, consider the generalized linear model --of which the Gaussian linear model, logistic/probit regression, count models, survival models (exponential and Weibull, not Cox), and many more commonly used models are special cases. In order to keep the notation general, we can write any likelihood function as 
\[
L (\bt | \bm{y}, \bm{X}) = \prod_{i=1}^{n} f ( y_{ij} | \bm{x}_{ij}, \bt)
\]
where $\bt$ is the vector of parameters to be estimated, $\bm{y}$ is the response vector (i.e., the dependent variable), $\bm{X}$ is the data matrix (i.e., the covariates), $f$ is some probability density function (e.g. Gaussian for a linear model, Poisson for a Poisson model, Bernoulli for logit/probit), $y_{ij}$ is the dyadic response between states $i$ and $j$, and $\bm{x}_{ij}$ is a vector of predictors measured on the $ij$ dyad. The very reason we are able to evaluate a joint likelihood for all the data by taking the product over dyadic probabilities is because of the assumption that all observations are independent and identically distributed (iid) conditional on the covariates. The ``identically distributed'' part of this assumption is less relevant to this discussion, but the ``independent'' part is central. Note that the structure of this model implies that dyadic outcomes used as the dependent variable in one observation cannot be used to construct independent variables in another observation. Such a specification, examples of which are proposed in \cite{Neumayer2010}, violate the independence assumption by construction. We have already considered the multitude of unreasonable and untenable theoretical assumptions imposed on the analyst through the dyadic design's necessity for iid, but now we see that we cannot even build a proper likelihood function absent this statistical independence. In other words, one can compute the product over the dyadic probabilities -- thus invoking the iid assumption, doing so will not produce errors or warnings in software -- but the result will be coefficients and standard errors with an arbitrary amount of bias. Indeed, the idea that one could produce an unbiased result with such a model violates the basic axioms of probability. 

Under what circumstances would a regression based on dyadic data be unbiased? Naturally, the model results will be unbiased when iid is satisfied. This implies all of the assumptions discussed in our theory section above. In the specific case in which dyadic events/connections do not depend on each other, and when subnational/national/k-adic/system effects are orthogonal to the dyadic predictors, no bias will be induced into the statistical model through application of a dyadic research design. Bias will necessarily be introduced in all other cases. Crucially, \emph{even if the researcher is only interested in the effect of one or more dyadic covariates (e.g., one has a genuinely dyadic theory about the effect of geographic distance on the likelihood of conflict), a dyadic design is still biased if there are unmodeled hyperdyadic dependencies that generated the data} -- as we demonstrate below with a simulation experiment. 

How severe will the bias in results from a dyadic design be, given that bias will be present in (nearly) all realistic data-analysis situations? This will depend on the specific case. Bias will range from minor, not affecting any substantive inferences, to severe, radically altering substantive inferences, depending on the extent to which un-modeled (in a dyadic design, un-modelable) dependencies are present. Generally speaking, the greater the un-modeled dependence, the greater the bias, but some models and functional forms are especially susceptible to this form of bias. It is often difficult to know, \emph{a priori}, the extent of these dependencies and thus the degree of bias that will be induced by following a dyadic design. One must also further wonder why a dyadic design would be applied when powerful techniques, such as the exponential random graph model (ERGM), are capable of modeling relational data whilst unifying levels of analysis and producing unbiased results \citep{Cranmer2011a}. 

ERGMs are statistical models for matrixes of relationships among actors (i.e., adjacency matrices in which the $ij$ cell records the 0/1 dyadic relationship between $i$ and $j$). Terms used to explain relationships in ERGMs can include dyadic covariates, as in dyadic regression, as well as higher order terms such as the number of reciprocated ties in the matrix (i.e., instances in which the $ij$ and $ji$ elements of the matrix are both 1) and the number of triangles in the matrix (i.e., the number of triples $\{i,j,k\}$ in which there is a tie between each pair in the triple). Though unlikely, it is possible that the network under study exhibits little or no hyperdyadic dependence, in which case an inferential network model may represent over-specification. In this case, where an independent dyadic design would be appropriate, the ERGM will reduce to logistic regression \citep{Cranmer2011a} and accurately reflect the lack of interdependence.   

By using inferential network analysis to study systems constituted by dyads, researchers can both account for hyperdyadic dependence and test whether a dyadic independence approach is appropriate. Further, it is important to be clear that taking a network inference approach does not imply complete interdependence among observations. Rather, network methods permit specification of and testing for precise forms of interdependence; the models account for as much interdependence as the analyst chooses to model. For example, if an ERGM is specified such that the terms include the number of mutual dyads in the network (i.e., to model reciprocity) and a list of dyadic covariates, then $y_{ij}$ only depends on $y_{ji}$ and is independent of all other potential relationships in the network.

Having discussed the problems inherent to dyadic regression and presented the ERGM as an archetypal solution to those problems, we should review two conditions such that if both conditions apply dyadic regression can be used in the context of hyperdyadic dependence. The first condition is that the researcher observes the timing of relationship states (i.e., the dyadic data is time-stamped). The second condition is that the state of relationships in period $t$ depends only upon the structure of relationships in time periods prior to $t$. In other words, there is no simultaneous dependence among relationships. If these two conditions apply, dyadic regression in which functions of the network of relationships in previous time periods (e.g., the number of shared partners between two states in the previous time point) operate as covariates to account for hyperdyadic dependence. In the context of temporal exponential random graph models (TERGMs), \cite{Hanneke2010} show that logistic regression can be used to estimate ``separable'' TERGMs, which are models in which the states of dyadic relationships at time $t$ are independent conditional upon the past states of the network.

 The danger of faulty inference is raised all the more by the fact that this type of bias will often falsely attributed explanatory power caused by the dependencies between dyadic connections to the covariates. We illustrate this attribution error below. As a result, we believe that the application of dyadic design by an analyst gives the reader substantial grounds upon which to doubt the validity of the study, both theoretically and empirically. We end our methodological discussion on a clear statement on the degree of hyperdyadic dependence necessary to justify a rejection of strictly dyadic methods. Dyadic designs should only be used if the researcher is comfortable assuming that there are no hyperdyadic dependencies. Otherwise there is no way to know {\em a priori} how results from strictly dyadic methods will differ from those that account for hyperdyadic dependence. This is not to say that all inferences made based on dyadic design, which encompasses most of the last 30 years of empirical research in international relations, are incorrect. We suspect that many are probably correct. But the above does suggest that most of this literature presents biased results and, absent an \emph{a priori} reason to believe that a variable of interest is orthogonal to the networks in which the observations occur or a direct replication that accounts for network dependencies, the inferences might reasonably be suspect. 

\subsection*{\centering{\textnormal{\textit{A Concise Illustration of the Consequences of Model Misspecification}}}}

The central focus of this essay has been a critique of approaches to studying pairwise state relationships in IR that apply assumptions of dyadic independence. Here, we provide a concise example that illustrates how the statistical argument for using multiple, rather than simple (one predictor), regression, seamlessly applies to the ways in which an {\em a priori} commitment to an approach that requires dyadic independence can bias inferences. Perhaps the strongest argument for a multiple regression approach is that it can adjust for the effects of confounders in estimating the relationship between one or more focal predictors and the dependent variable. In exactly the same way, higher order network structures,  for which it is impossible to adjust in dyadic independence models, can confound relationships between covariates and dyadic outcome variables. One of the major threats to inference presented by confounders is that failure to adjust for them can result in a considerably increased Type 1 error rate in hypothesis testing. That is, failure to adjust for confounders can lead researchers to reject the null hypothesis of no relationship when there is actually no relationship at a higher rate than they would if they were to adjust for the confounders. Here, we show how dyadic regression in the presence of confounding hyperdyadic dependence can lead to a dramatically inflated Type 1 error rate. 

We use the Exponential Random Graph family of models (i.e., the ERGM) \citep{Holland1981,Wasserman1996,Cranmer2011a} to generate relational data with a hyperdyadic dependence structure. Using the ERGM, we generate random networks in which the tie $Y_{ij}$ depends upon all other directed dyads in which $j$ is the recipient, which we denote $Y_{-i,j}$. The form of the exponential random graph model that we use to generate networks is given by the following specification:
$$ P(Y) = \frac{\exp\{ -3.5\times E(Y) +0.75\times IS_2(Y) -0.1\times IS_3(Y) \}}{C(\theta)},$$
where $E(Y)$ is the number of edges in the network, $IS_2(Y)$ is the number of in-two-starts in the network and $IS_3(Y)$ is the number of in-three-stars in the network. $C(\theta)$ is the normalizing constant. For readers interested in more detail on the  ERGM and ERGM specification see e.g.,  \citet{Wasserman1996}, \citet{Cranmer2011a}, \citet{Desmarais:2012physa}, and \citet{Desmarais2012psj}. The edges term, which counts the number of connections in the network, is the ERGM equivalent of an intercept term in regression analysis. An in-two-star is a configuration in which two nodes (states in this case) send a tie to a third node. For each triple of nodes in a directed network there are three possible in-two-stars (i.e., one in which each of the nodes assumes the role of recipient of the two ties that form the in-two-star). When there is a positive parameter associated with in-two-stars in an ERGM, senders of new ties will seek to send ties to nodes that already receive many ties (i.e., a popularity effect). An in-three-star, as the name implies, is a configuration in which three nodes each send a tie to a target node. If in-three-stars is included in ERGM along with in-two-stars, and exhibits a negative parameter value, the popularity effect lessens as a target node gains more ties. That is, with positive in-two-stars and negative in-three-stars, a node becomes a more attractive recipient as it begins to receive ties, but attraction to a node lessens as it gains a large number of ties (i.e., analogous to a quadratic relationship in regression). This specification is chosen to give a simple example of hyperdyadic interdependences and the parameter values are selected to create a generative process based on moderated popularity.  We generate 500 directed networks with 25 nodes from this ERGM. 

We introduce potential confounding bias that is related to hyperdyadic dependence by constructing a covariate that derives from network structure, and is related to the number of ties sent to the two nodes in a dyad. Suppose the relationships that constitute the simulated networks are the issuance of economic sanctions such that $Y_{ij}=1$ indicates that $i$ has issued an economic sanction against $j$. Now suppose that $X_{ij}$ represents the degree to which state $i$'s interests overlaps with $j$'s (e.g., as measured through voting in the United Nations General Assembly). Further assume that two states are likely to find their interests aligned when they are sanctioned by the same states (i.e., the enemy of an enemy is a friend). Thus, the number of third states that sanction both $i$ and $j$ is a positive causal predictor of $X_{ij}$, the interest alignment between $i$ and $j$. Note that $i$ and $j$ are likely to have common sanctioners if $i$ and $j$ are popular in the sanctions network. Because the generative model for the artificial sanctions network is based on popularity, any variable that is caused by the popularity of nodes in the sanction network will be spuriously correlated with dyadic tie formation in the sanctions network, especially if the empirical model does not account for the process of tie-formation based on popularity. We construct  $X_{ij}$ as the number of nodes in the simulated network that send ties to both $i$ and $j$, plus a normally distributed error. The confounding variable in this example is the number of nodes that send ties to both $i$ and $j$. Via the popularity dynamics built into the ERGM, this variable is related to whether $i$ sends a tie to $j$ (i.e., if many nodes send a tie to $j$, then $i$ will want to send ties to $j$ {\em and} $i$ will have senders in common with $j$ simply due to the large number of nodes sending to $j$). This variable is also related to $X_{ij}$, our hypothetical interest overlap variable, in that states' interests align when they are sanctioned by common states. If a model included interest overlap between $i$ and $i$ (i.e., $X_{ij}$ as a dyadic predictor of whether $i$ sanctioned $j$ (i.e., $Y_{ij}$) without accounting for the popularity of both $i$ and $j$ among other states, there would be a high risk of incorrectly rejecting the null hypothesis that interest overlap between $i$ and $j$ does not cause $i$ to sanction $j$.

\begin{figure}[ht]
\begin{center}
\includegraphics[scale=1.4]{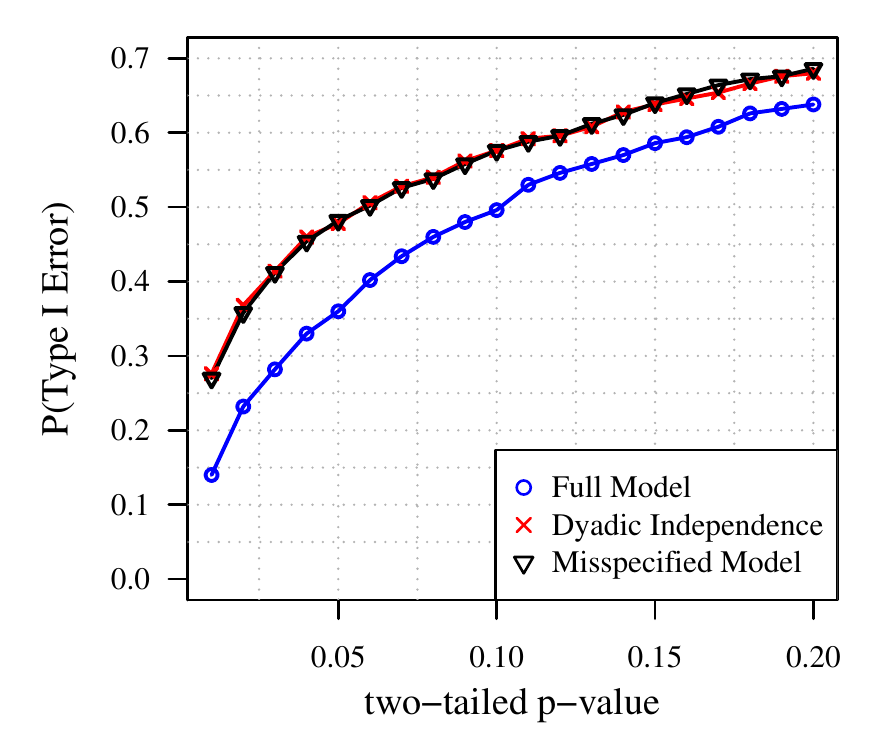}~~~~~~~
\end{center}
\vspace{-1cm}
\caption{Type 1 error rates for different specifications of an exponential random graph model. The dyadic independence model is logistic regression. Error rates are calculated over 500 iterations of simulations of a 25 node network from an exponential random graph.} 
\label{fig:error}
\end{figure}

We estimated three different statistical models in each iteration of our simulation study. The first was a dyadic logistic regression in which $Y_{ij}$ was regressed on $X_{ij}$ (Dyadic Independence). The second was a model in which $X_{ij}$ was added to the correctly specified ERGM (Full Model). And the third was a model in which $X_{ij}$ was included in an erroneously specified ERGM (Misspecified Model). \footnote{\doublespacing The interdependence terms included in the Misspecified model include the geometrically weighted edgewise shared partners statistic (GWESP) and the geometrically weighted out degree statistic (GWODEGREE). GWESP models transitivity in a network -- the tendency for two nodes to form a tie if those nodes have one or more common (i.e., shared) partners. GWODEGREE models varying levels of sending (i.e., out ties) by nodes. See \cite{Snijders2006} for detailed discussion of these statistics. }  In Figure \ref{fig:error}, we present Type 1 error rates generated in our simulations. We see that the full model exhibits a Type 1 error rate that is approximately 10--15 percentage points lower than that of dyadic regression or a misspecified ERGM at conventionally used levels of statistical significance. This simple simulation exercise illustrates that the omission of hyper-dyadic structure from a model of relational data can increase Type 1 error through confounding. Note that this is exactly what will happen with omitted variable bias. To wit, the inferential pitfalls associated with omitting hyper-dyadic relational structure from a model of dyadic data are indistinguishable from those of the omitted variable bias. The \R  code needed to re-produce this simulation is given in the Appendix.

\section*{\centering{Conclusion}}

It is not valid to use dyadic methods in the face of hyperdyadic generative processes to draw model-based inferences. Model based inferences are only as accurate as the statistical model specification, and to relegate one's specification to the class of dyadic independence models is to erect an artificial barrier to sound model specification and inferenece. A researcher may tolerate the use of an inferior class of statistical models in order to ease computational burdens, but such tolerance is a purely individual and subjective question, which is beyond the focus of our current commentary. Dyadic design is widespread, but we cannot quantify the extent to which false inferences have been made without individually replicating each analysis; we can be mathematically confident that all/most include some level of bias. The question then becomes how much stock should we put in results that are known to be biased? We do not believe that acknowledging the bias to readers and moving on is sufficient because the degree of bias is unquantified. Just like in a 12-step program, admitting you have a problem is important, but is not enough on its own. 

We argued that dyadic design is problematic from both a theoretical and empirical perspective. To summarize, we can only expect unbiased statistical results in extremely unlikely situations, such as dyads not depending on one another in any way beyond the conditioning on relational measures within the dyad, and even in such situations, dyadic design imposes a strong and usually substantively untenable set of assumptions. We believe that these assumptions will rarely if ever be met in interesting international politics applications, and, when they do not cause bias or model misspecification, these assumptions result in treatments of the data that are wildly unrealistic to even the analyst with the most basic knowledge of geopolitics. 

Where does this leave us? We are of the opinion, and hope that the reader will also have been convinced by this discussion, that dyadic design is nearly never appropriate for international relations theory or data. The critical reader may convince herself that one or even several of the points mentioned above do not apply to her data or problem, but we would have to be outright wrong about \emph{every point} we made above in order for dyadic design to be superior to other easily available approaches. We see reason to doubt many published results based on dyadic design, we do not believe that our critique should be cause for despair in the community of international relations scholars. Rather, the development of empirical tools that can account for nearly arbitrarily complex forms of interdependence present the opportunity for great strides to be made in international relations theory. Our previous work in this area would suggest, for example, that triadic closure (ties between all dyads in the $ijk$ triple) is very common in positively valanced (e.g., a tie is a ``good'' thing) networks like alliances \citep{Cranmer2012a,Cranmer2012b} and very uncommon in negatively valanced networks like conflict \citep{Cranmer2011a}. A wide variety of similar processes may be at work in international phenomena and are waiting to be discovered. We believe that the field of international relations is on the cusp of a revolution in how we treat complexity, both theoretically and empirically, and we look forward to an exciting period of innovation as such complexity-oriented approaches permeate the field.

\section*{\centering{Appendix}}

\singlespacing
\begin{verbatim}
# Set directory in which to store plots and results
# Users should change this to the full path to folder on the local machine
setwd("~/Desktop/")

# set the seed in order to replicate exactly
set.seed(123)

# Necessary libraries
# sna is a general network analysis package with many useful functions
library(sna)
# ergm is a package for exponential random graph model estimation
library(ergm)

# set the number of iterations in the simulation
# increase for greater accuracy in error rate estimates
nsim <- 500

# create vectors in which to store p-values for each model
null.p <- numeric(nsim)
omit.p <- numeric(nsim)
miss.p <- numeric(nsim)

# loop over nsim iterations
for(i in 1:nsim){
	# function rgraph in sna creates i.i.d. Bernoulli networks, here with probability 0.1
	# net will serve at the initial network for simulating from ERGM using MCMC
	net <- rgraph(25,1,.1)
	# the simulate function simulates from ERGM
	# MCMC starts at net and converges to a hyperdyadic dependence network
	# edges term of -3.5 leads to relatively sparse networks (like most networks in IR)
	# istar(2) term of 0.75 is a relatively strong popularity effect
	# istar(3) term of -0.1 serves to moderate the popularity effect
	net0 <- simulate(net~edges+istar(2)+istar(3), coef= c(-3.5,.75,-.1))
	# dyadX is the shared sender spurious covariate
	dyadX <- t(net0[,])%*%net0[,]+matrix(rnorm(25*25,sd=3/4),25,25)
	# Run the correctly specified model
	null_est <- ergm(net0 ~ edges+edgecov(dyadX)+istar(2:3))
	# Run the misspecified erg
	miss_est <- ergm(net0 ~ edges+edgecov(dyadX)+gwesp(0,fixed=T)+gwodegree(1,fixed=T))
	# Run dyadic logit
	omit_est <- ergm(net0 ~ edges+edgecov(dyadX))
	
	# Extract and store p-values
	null.p[i] <- summary(null_est)$coefs[2,4]
	omit.p[i] <- summary(omit_est)$coefs[2,4]
	miss.p[i] <- summary(miss_est)$coefs[2,4]
	
	# print iteration to keep track/time
	print(i)
	
}

# save results
save(list=c("null.p","omit.p","miss.p"),file="SimulationResults.RData")

# Vector of significance levels at which to calculate the Type I error rate
p.vals <- seq(0.01,0.2,by=0.01)

# Calculate the Type I error rate of each model at each p-value
null.error <- numeric(length(p.vals))
omit.error <- numeric(length(p.vals))
miss.error <- numeric(length(p.vals))
for(i in 1:length(p.vals)){
	null.error[i] <- mean(null.p < p.vals[i])
	omit.error[i] <- mean(omit.p < p.vals[i])
	miss.error[i] <- mean(miss.p < p.vals[i])
}

# Make the results plot
filei <- "type_one_error.pdf"
pdf(filei,height=3,width=3.5,pointsize=9,family="Times")
par(las=1,mar=c(4,5,1,1),cex.lab=1.25,cex.axis=1)
plot(p.vals,null.error,type="n",ylab="P(Type I Error)",xlab="",ylim=c(0,0.7))
abline(h=seq(.05,.7,by=.05),lty=3,col="grey70")
abline(v=seq(.025,2,by=.025),lty=3,col="grey70")
lines(p.vals,omit.error,lwd=1.5,col="red")
lines(p.vals,null.error,lwd=1.5,col="blue")
lines(p.vals,miss.error,lwd=1.5)
points(p.vals,omit.error,lwd=1.5,col="red",pch=4,cex=.85)
points(p.vals,null.error,lwd=1.5,col="blue",pch=1,cex=.85)
points(p.vals,miss.error,lwd=1.5,pch=6,cex=.85)
title(xlab="two-tailed p-value",line=2.25)
#abline(0,1)
legend("bottomright",legend=c("miss.pecified Model","Dyadic Independence","Full Model")[c(3,2,1)],col=c("black","red","blue")[c(3,2,1)],pch=c(6,4,1)[c(3,2,1)],bg="white")
dev.off()


\end{verbatim}

\clearpage
\bibliographystyle{apa}
\bibliography{dyad}

\begin{thebibliography}{}

\bibitem[\protect\astroncite{Bennett and Stam}{2004}]{Bennett2004}
Bennett, D.~S. and Stam, A.~C. (2004).
\newblock {\em The Behavioral Origins of War}.
\newblock University of Michigan Press, Ann Arbor.

\bibitem[\protect\astroncite{Bremer}{1992}]{Bremer1992}
Bremer, S. (1992).
\newblock Dangerous dyads: Conditions affecting the likelihood of interstate
  war, 1816-1965.
\newblock {\em Journal of Conflict Resolution}, 36(2):309--341.

\bibitem[\protect\astroncite{Bremer et~al.}{2003}]{BremerReganClark2003}
Bremer, S., Regan, P.~M., and Clark, D.~H. (2003).
\newblock Building a science of world politics.
\newblock {\em Journal of Conflict Resolution}, 47:3--13.

\bibitem[\protect\astroncite{Bueno~de Mesquita and
  Lalman}{1992}]{BdmLalman1992}
Bueno~de Mesquita, B. and Lalman, D. (1992).
\newblock {\em War and Reason}.
\newblock Yale University Press, New Haven, CT.

\bibitem[\protect\astroncite{Chen}{2010}]{Chen2010}
Chen, H.~Y. (2010).
\newblock Compatibility of conditionally specified models.
\newblock {\em Statistics \& probability letters}, 80(7):670--677.

\bibitem[\protect\astroncite{Cranmer and Desmarais}{2011}]{Cranmer2011a}
Cranmer, S.~J. and Desmarais, Bruce, A. (2011).
\newblock Inferential network analysis with exponential random graph models.
\newblock {\em Political Analysis}, 19(1):66--86.

\bibitem[\protect\astroncite{Cranmer et~al.}{2012a}]{Cranmer2012b}
Cranmer, S.~J., Desmarais, B.~A., and Kirkland, J.~H. (2012a).
\newblock Towards a network theory of alliance formation.
\newblock {\em International Interactions}, 38(3):295--324.

\bibitem[\protect\astroncite{Cranmer et~al.}{2012b}]{Cranmer2012a}
Cranmer, S.~J., Desmarais, B.~A., and Menninga, E.~J. (2012b).
\newblock Complex dependencies in the alliance network.
\newblock {\em Conflict Management and Peace Science}, 29(3):279--313.

\bibitem[\protect\astroncite{Cranmer et~al.}{2014}]{Cranmer2014son}
Cranmer, S.~J., Heinrich, T., and Desmarais, B.~A. ({2014}).
\newblock {Reciprocity and the Structural Determinants of the International
  Sanctions Network}.
\newblock {\em {Social Networks}}, {36}({January}):{5--22}.

\bibitem[\protect\astroncite{Desmarais and Cranmer}{2012a}]{Desmarais2012psj}
Desmarais, B.~A. and Cranmer, S.~J. ({2012}a).
\newblock Micro-level interpretation of exponential random graph models with
  application to estuary networks.
\newblock {\em Policy Studies Journal}, 40(3):402--434.

\bibitem[\protect\astroncite{Desmarais and
  Cranmer}{2012b}]{Desmarais:2012physa}
Desmarais, B.~A. and Cranmer, S.~J. (2012b).
\newblock Statistical mechanics of networks: Estimation and uncertainty.
\newblock {\em Physica A: Statistical Mechanics and its Applications},
  391(4):1865 -- 1876.

\bibitem[\protect\astroncite{Dorff and Ward}{2013}]{Dorff2013}
Dorff, C. and Ward, M.~D. (2013).
\newblock Networks, dyads, and the social relations model.
\newblock {\em Political Science Research and Methods}, 1(02):159--178.

\bibitem[\protect\astroncite{Ellis et~al.}{2010}]{Ellis2010}
Ellis, G., Mitchell, S.~M., and Prins, B.~C. (2010).
\newblock How democracies keep the peace: Contextual factors that influence
  conflict management strategies.
\newblock {\em Foreign Policy Analysis}, 6(4):373--398.

\bibitem[\protect\astroncite{Hanneke et~al.}{2010}]{Hanneke2010}
Hanneke, S., Fu, W., Xing, E.~P., et~al. (2010).
\newblock Discrete temporal models of social networks.
\newblock {\em Electronic Journal of Statistics}, 4:585--605.

\bibitem[\protect\astroncite{Holland and Leinhardt}{1981}]{Holland1981}
Holland, P.~W. and Leinhardt, S. (1981).
\newblock An exponential family of probability distributions for directed
  graphs.
\newblock {\em Journal of the American Statistical Association},
  76(373):33--50.

\bibitem[\protect\astroncite{Manger et~al.}{2012}]{manger2012hierarchy}
Manger, M.~S., Pickup, M.~A., and Snijders, T.~A. (2012).
\newblock A hierarchy of preferences: A longitudinal network analysis approach
  to pta formation.
\newblock {\em Journal of Conflict Resolution}, 56(5):853--878.

\bibitem[\protect\astroncite{Maoz and Russett}{1993}]{MaozRussett1993}
Maoz, Z. and Russett, B. (1993).
\newblock Normative and structural causes of democratic peace.
\newblock {\em American Political Science Review}, 87(3):624--638.

\bibitem[\protect\astroncite{Mitchell et~al.}{1999}]{Mitchell1999}
Mitchell, S.~M., Gates, S., and Hegre, H. (1999).
\newblock Evolution in democracy-war dynamics.
\newblock {\em Journal of Conflict Resolution}, 43(6):771--792.

\bibitem[\protect\astroncite{Neumayer and Pl\"{u}mper}{2010}]{Neumayer2010}
Neumayer, E. and Pl\"{u}mper, T. (2010).
\newblock Spatial effects in dyadic data.
\newblock {\em International Organization}, 64:145--166.

\bibitem[\protect\astroncite{Oneal and Russett}{1999}]{Oneal1999}
Oneal, J.~R. and Russett, B. (1999).
\newblock The kantian peace: The pacific benefits of democracy,
  interdependence, and international organizations, 1885-1992.
\newblock {\em World Politics}, 52(1):1--37.

\bibitem[\protect\astroncite{Poast}{2010}]{Poast:2009}
Poast, P. (2010).
\newblock (mis)using dyadic data to analyze multilateral events: An application
  to alliance formation.
\newblock {\em Political Analysis}, 18(4).

\bibitem[\protect\astroncite{Snijders et~al.}{2006}]{Snijders2006}
Snijders, T.~A., Pattison, P.~E., Robins, G.~L., and Handcock, M.~S. (2006).
\newblock New specifications for exponential random graph models.
\newblock {\em Sociological methodology}, 36(1):99--153.

\bibitem[\protect\astroncite{Tilly}{2003}]{tilly2003politics}
Tilly, C. (2003).
\newblock {\em The politics of collective violence}.
\newblock Cambridge University Press.

\bibitem[\protect\astroncite{Ward and Hoff}{2007}]{Ward:2007}
Ward, M.~D. and Hoff, P.~D. (2007).
\newblock Persistent patterns of international commerce.
\newblock {\em Journal of Peace Research}, 44(2):157--175.

\bibitem[\protect\astroncite{Ward et~al.}{2011}]{Ward2011}
Ward, M.~D., Stovel, K., and Sacks, A. (2011).
\newblock Network analysis and political science.
\newblock {\em Annual Review of Political Science}, 14:245--264.

\bibitem[\protect\astroncite{Warren}{2010}]{Warren2010}
Warren, C. (2010).
\newblock The geometry of security: Modeling interstate alliances as evolving
  networks.
\newblock {\em Journal of Peace Research}, 47(6):697--709.

\bibitem[\protect\astroncite{Wasserman and Pattison}{1996}]{Wasserman1996}
Wasserman, S. and Pattison, P. (1996).
\newblock Logit models and logistic regressions for social networks: I. an
  introduction to markov graphs and $p^*$.
\newblock {\em Psychometrika}, 61(3):401--425.

\bibitem[\protect\astroncite{White}{2008}]{white2008identity}
White, H.~C. (2008).
\newblock {\em Identity and control: How social formations emerge}.
\newblock Princeton University Press.

\end{thebibliography}

\end{document}